\documentclass[showpacs,showkeys,superscriptaddress,aps]{revtex4}
\def\be{\begin{eqnarray*}}
\def\ee{\end{eqnarray*}}
\begin{document}
%%%%%%%%%%%%%%%%%%%%%%%%%%%%
%%%%%%%%%%%%%%%%%%%%%%%%%%%%
\title{A Realization of Matrix KP Hierarchy by Coincident D-brane States}
%%%%%%%%%%%%%%%%%%%%%%%%%%%%%%%%%%%%%%%
\author{Hironori YAMAGUCHI}
\email[email : ]{hironori@kiso.phys.metro-u.ac.jp}
\affiliation{Department of Physics, Tokyo Metropolitan University,\\
Minamiohsawa 1-1, Hachiohji, Tokyo, 192-0397 Japan}
%%%%%%%%%%%%%%%%%%%%%%%%%%%%%%%%%%%%%%%
\author{Satoru SAITO}
\email[email : ]{saito_ru@nifty.com}
\affiliation{Hakusan 4-19-10, Midoriku, Yokohama 226-0006 Japan}
%%%%%%%%%%%%%%%%%%%%%%%%%%%%%%%%%%%%%%%
%%%%%%%%%%%%%%%%%%%%%%%%%%%%
\begin{abstract}
We describe the Sato-Wilson type formulation of the KP hierarchy within the framework of closed string theory. A matrix generalization of this formalism is shown to allow natural interpretation of coincident D-branes as a sourse of nonabelian gauge theory.
\end{abstract}
%%%%%%%%%%%%%%%%%%%%%%%%%%%%
\maketitle

%%%%%%%%%%%%%%%%%%%%%%%%%%%%%%%%%%%%%%%%
\pagestyle{plain}
\def\be{\begin{eqnarray*}}
\def\ee{\end{eqnarray*}}
\def\ve{\vfill\eject}
\newcommand{\bmm}[1]{\mbox{\boldmath$#1$}}
%%%%%%%%%%%%%%%%%%%%%%%%%%%%%%%%%%%%%%%%%%%
%%%%%%%%%%%%%%%%%%%%%%%%%%%%%%%%%%%%%%%%%%
\section{Introduction}

D-branes constitute an important part of the superstring theory in its nonperturbative description. Especially coincident multi D-brane states enable us to reproduce nonabelian guage theory in the low energy. From the mathematical point of view, on the other hand, a non-perturbative field theory must be formulated within the framework of a completely integrable system, namely a system whose whole possible solutions can be determined analytically. If a system is nonintegrable there remain some domains which refuse our nonperturbative analysis in principle. Therefore it is natural to ask if there exists a way to describe a D-brane dynamics in terms of integrable systems.

The correspondence between open string correlation functions and the KP hierarchy of soliton theory has been known for a long time\cite{S}. Recently we have extended\cite{SS1} the relation to the case of closed strings and shown that a boundary state of conformal strings satisfies Hirota-Miwa equation. Moreover we generalized Hirota-Miwa equation in order to characterize dynamics of D-branes on which tachyon fields are condensed\cite{SS2}. 

We would like to extend our argument further, in this paper, to establish a correspondence between multi D-brane states and integrable systems. In particular we will show that a multi D-brane state satisfies the matrix generalization of the KP hierarchy, thus leading to a natural interpretation of coincident D-branes as a sourse of non-abelian gauge theory.

%%%%%%%%%%%%%%%%%%%%%%%%%%%%%%%%%%%%%
\section{A description of Sato-Wilson formalism in terms of closed string theory}

To begin with we review briefly the Sato-Wilson formalism of the KP hierarchy\cite{Sato, SW}, in a way convenient for our discussion providing a simple generalization to a matrix form. Starting from a simple function
\begin{equation}
\Psi_0=e^{\xi(z)},\qquad \xi(z)=-\sum_{n=0}^\infty t_nz^{n},
\label{Psi_0}
\end{equation}
with $t=\{t_n\}$ being the set of time variables characterizing infinitely many soliton equations in the KP hierarchy, M.Sato has shown\cite{Sato} that general solution to the eigenvalue problems
\begin{equation}
L\Psi=z\Psi,\quad {\partial\Psi\over\partial t_n}=B_n\Psi,\quad n=0,1,2,...
\label{Lax eq}
\end{equation}
, for given Lax pairs $L$ and $B_n$, is given by
\begin{equation}
\Psi(z)=W\Psi_0(z).
\label{Psi=WPsi_0}
\end{equation}
Here the pseudo differential operator $W$
\begin{equation}
W=\sum_{n=0}^\infty W_n{\partial^{-n}\over\partial t_1^{-n}}
\label{W}
\end{equation}
, which is also called Sato-Wilson operator, is defined from $L$ through $L=W{\partial\over \partial t_1}W^{-1}$. The consistency relation of (\ref{Lax eq}) leads to infinitely many soliton equations among $W_n$'s, {\it i.e.}, the KP hierarchy. We can also prove the equivalence of this scheme to the single equation
\begin{equation}
\oint {dz\over 2\pi i}\Psi(t',z)\Psi^*(t,z)=0,
\label{oint dzPsi(t,z)Psi^(t',z)=0}
\end{equation}
in which the integration contour is assumed to encircle the origin $z=0$. The well known Hirota-Miwa equation \cite{Hirota}\cite{Miwa} can be derived from this expression\cite{DJKM}.

An important ingredient of this simple formulation of the KP-hierarchy is the fact that $\xi(z)$ is expanded into the positive powers of $z$ while $W(z)$ is into the negative powers of $z$. It seems, at a glance, rather difficult to relate such formalism with the theory of closed strings, because the string coordinate $X(z,\bar z)$ includes both $X(z)$ and $\tilde X(\bar z)$ which inevitably mix $z$ and $z^{-1}$. We will solve, however, this problem by choosing proper combination of modes of the string coordinates. This solution will be seen not only overcome the problem but also leads to a natural interpertation of coincident D-branes.

We are now going to translate this formalism into the language of closed string theory. In particular we are interested in the nature of world sheets attached on D-branes. To clarify the role of duality we introduce an index $\zeta$ and distinguish the closed string coordinate whether it is in the ordinary theory ($\zeta=1$) or in the T-dual theory ($\zeta=-1$).  We also suppress the space-time indices in this paper. Hence the space-time coordinates of bosonic and fermionic closed strings are written as 
$$
X^\zeta(z,\bar z)=X(z)+\zeta\tilde X(\bar z),\qquad
\psi^\zeta(z,\bar z)=\psi(z)+i\zeta\tilde \psi(\bar z).
$$
Although the original Sato-Wilson formalism has nothing to do with supersymmetry\cite{UY} we discuss the fermionic component of the strings parallel with bosonic one, so that we will find supersymmetric version of the correspondence between the KP hierarchy and the string theory.

When we discuss string theory we parameterise the world sheet by $(\sigma, \tau)$ related to $z$ through $z=e^{-i\sigma+\tau}$. The D-brane, on which the strings are attached, is placed at $\tau=0$, so that the fields are expanded according to
\begin{eqnarray}
X^\zeta(\sigma)&=&X_<^\zeta(\sigma)+X_>^\zeta(\sigma),\qquad
X_{\stackrel{<}{>}}^\zeta(\sigma)
=
\left(\matrix{x_0\cr 0\cr}\right)\pm i\sum_{n=1}^\infty{1\over n}\left(\alpha_{\pm n}-\zeta\tilde\alpha_{\mp n}\right)e^{\pm in\sigma},
\label{X<>}\\
\psi^\zeta(\sigma)&=&e^{i\sigma/2}\left(\psi^\zeta_<(\sigma)+\psi^\zeta_>(\sigma)\right),\qquad
\psi^\zeta_{\stackrel{<}{>}}(\sigma)=\sum_{r={1\over 2}}^\infty \left(\psi_{\pm r}+i\zeta\tilde\psi_{\mp r}\right)e^{\pm ir\sigma}.
\end{eqnarray}
The bare boundary state is given by
\begin{equation}
|\rho,\eta\rangle=\prod_{n=1,r={1\over 2}}^\infty\exp\left[{\rho\over n}\alpha_{-n}\tilde\alpha_{-n} +i\eta\rho \psi_{-r}\tilde\psi_{-r}\right]|0\rangle
\label{R_b}
\end{equation}
with $\eta$ being the spin structure and $\rho=1,0,-1$ corresponding the Dirichlet, free and the Neumann boundary conditions, respectively. 

The significance of the special combinations of the components in $X_{\stackrel{<}{>}}^\zeta(\sigma)$ and $\psi^\zeta_{\stackrel{<}{>}}(\sigma)$ becomes clear if we operate them to the boundary state:
\begin{eqnarray*}
X_{\stackrel{<}{>}}^\zeta(\sigma)|\rho,\eta\rangle=0,\qquad 
\langle\rho,\eta|X_{\stackrel{<}{>}}^\zeta(\sigma)=0,\qquad &&{\rm if}\quad \zeta\rho=1,\\
\psi_{\stackrel{<}{>}}^\zeta(\sigma)|\rho,\eta\rangle=0,\qquad 
\langle\rho,\eta|\psi_{\stackrel{<}{>}}^\zeta(\sigma)=0,\qquad &&{\rm if}\quad \zeta\rho\eta=-1.
\end{eqnarray*}

Having prepared these tools we can proceed our argument and define the following operators
\begin{eqnarray}
\hat V^\zeta&=&\exp\left[{i\sqrt2}\sum_j\left(k_jX^\zeta_<(\sigma_j)-i\kappa_j\psi^\zeta_<(\sigma_j)\right)\right],
\label{hat V}\\
\hat W^\zeta
&=&\exp\left[{i\over 4\pi}\int_0^{2\pi}\left(A(\sigma){\partial X_>^\zeta(\sigma)\over\partial\sigma}+i \Theta(\sigma)\psi_>^\zeta(\sigma)e^{-i\sigma/2}\right)d\sigma\right],
\label{hat}
\end{eqnarray}
where $A(\sigma)$ and $\Theta(\sigma)$ are assumed being expanded as follows.
$$
A(\sigma)=\sum_{n=1}^\infty A_ne^{in\sigma},\qquad
\Theta(\sigma)=\sum_{r={1\over 2}}^\infty \Theta_r e^{ir\sigma+i\sigma/2}.
$$
We also introduce the vertex operator
\begin{equation}
\phi^\zeta(\sigma)=
:\exp\left[{i\over \sqrt 2}\left(X^\zeta(\sigma)+i\epsilon\psi^\zeta(\sigma)\right)\right]:
\label{phi(sigma)}
\end{equation}
by which we obtain, under the exchange with $\hat V^\zeta$ and $\hat W^\zeta$, the following useful formula:
\begin{eqnarray}
\hat V^\zeta\phi^{\zeta'}(\sigma)&=&e^{(1-\zeta\zeta')(\xi_b(\sigma)+i\epsilon \xi_f(\sigma))}\phi^{\zeta'}(\sigma)\hat V^\zeta,\label{Vphi}\\
\hat W^\zeta\phi^{\zeta'}(\sigma)&=&e^{(1-\zeta\zeta')(A(\sigma)+i\epsilon\Theta(\sigma))/(2\sqrt 2 i)}\phi^{\zeta'}(\sigma)\hat W^\zeta,\label{Wphi}
\end{eqnarray}
where
$$
\xi_b(\sigma)=-\sum_jk_j\left(i\sigma_j+\sum_{n=1}^\infty{1\over n} e^{in(\sigma_j-\sigma)}\right),
\qquad
\xi_f(\sigma)=i\sum_j\kappa_j\sum_{r={1\over 2}}^\infty e^{ir(\sigma_j-\sigma)+i\sigma/2}.
$$

We are now ready to write down the supersymmetric version of Sato-Wilson function $\Psi$ in terms of the string theory. By using (\ref{Vphi}) and (\ref{Wphi}) we calculate a correlation function as follows
\begin{equation}
\langle\rho',\eta'|\hat W^\zeta\hat V^\zeta\phi^{\zeta'}(\sigma)|\rho,\eta\rangle
=
e^{(1-\zeta\zeta')(A(\sigma)+i\epsilon\Theta(\sigma))/(2\sqrt 2 i)}e^{(1-\zeta\zeta')(\xi_b(\sigma)+i\epsilon\xi_f(\sigma))}
\langle\rho',\eta'|\phi^{\zeta'}(\sigma)\hat W^\zeta\hat V^\zeta|\rho,\eta\rangle.
\label{exchange}
\end{equation}
We notice that, when $\zeta'=-\zeta, \zeta\rho=\zeta'\rho'=1$ are satisfied, the factor $\langle\rho',\eta'|\phi^{\zeta'}(\sigma)\hat W^\zeta\hat V^\zeta|\rho,\eta\rangle$ in the right hand side of (\ref{exchange}) turns to $\langle\rho',\eta'|\rho,\eta\rangle$. Under this circumstance let us consider the case that there is no fermionic mode ({\it i.e.,} $\epsilon=0$). If we identify $t_n$'s in $\Psi_0$ of (\ref{Psi_0}) with
\begin{equation}
t_0=i\sum_jk_j\sigma_j,\quad t_n={1\over n}\sum_jk_je^{in\sigma_j}\qquad n=1,2,3,...
\label{Miwa transformation}
\end{equation}
and write $W(z)$ in $\Psi(z)$ of (\ref{Psi=WPsi_0}) with
$$
W(z)=e^{-iA(\sigma)/\sqrt 2},
$$
we find that
\begin{equation}
\Psi_0(z)={\langle\rho',\eta'|\hat V^\zeta\phi^{\zeta'}(\sigma)|\rho,\eta\rangle\over\langle\rho',\eta'|\rho,\eta\rangle},\qquad 
\Psi(z)={\langle\rho',\eta'|\hat W^\zeta\hat V^\zeta\phi^{\zeta'}(\sigma)|\rho,\eta\rangle\over\langle\rho',\eta'|\rho,\eta\rangle}.
\label{Psi(z)=string correlator}
\end{equation}
hold. In this way the Sato-Wilson functions are expressed in the form of correlation functions of closed bosonic strings propagating between two boundaries.

It is then quite natural to interprete (\ref{Psi(z)=string correlator}) as a supersymmetric generalization of the Sato-Wilson function when $\epsilon$ is recovered.

Some comments are in order.

1) The relation of $t_n$'s and $k_j$'s in (\ref{Miwa transformation}) is nothing but the Miwa transformation\cite{Miwa}.

2) The functions $W(z)$ and $\xi(z)$ in the right hand side of (\ref{Psi(z)=string correlator}) are expanded into powers of $z^{-1}=e^{i\sigma}$ and $z=e^{-i\sigma}$, respectively, as we expected.

3) The fermionic nature of the field $\phi^\zeta(z,\bar z)$, the analytic continuation of (\ref{phi(sigma)}) to the world sheets from the boundary, enables us to derive the supersymmetric KP hierarchy in the form
\begin{equation}
\oint {dz\over 2\pi i}\oint {d\bar z\over 2\pi i}\Psi(t',z,\bar z)\Psi^*(t,z,\bar z)=0
\label{oint dz oint dbar zPsi(t,z,bar z)Psi^(t',z,bar z)=0}
\end{equation}
by using the standard technique of universal Grassmannian\cite{SS2}.

%%%%%%%%%%%%%%%%%%%%%%%%%%%%%%%%%%%%%%

\section{Matrix generalization of the correspondence and D-brane interpretation}

In the recent paper \cite{SS2} a tachyon field was introduced from which the de Alwis type of boundary state\cite{de Alwis} was reproduced as a special case. Within the framework of our formulation in the present paper it corresponds to the state 
$$
|A,\Theta,\rho,\eta\rangle=\hat W^\zeta|\rho,\eta\rangle,\qquad \zeta\rho=-1,\ \eta=-1.
$$
If we write $A(\sigma)$ and $\Theta$ in $\hat W$ as
$$
A(\sigma)=-\int_0^{2\pi}{d\sigma'\over 2\pi}K(\sigma')\sum_{n=1}^\infty{c_n\over n}\left(e^{in(\sigma-\sigma')}+e^{-in(\sigma-\sigma')}\right),
$$
$$
\Theta(\sigma)=\int_0^{2\pi}{d\sigma'\over 2\pi}{\cal K}(\sigma')\sum_{r=1}^\infty c_r\left(e^{in(\sigma-\sigma')}+e^{-in(\sigma-\sigma')}\right)
$$
and set
$$
c_n=\left(1+\left({u\over n}\right)\right)^{-1/2}, \qquad  c_r=\left(1+\left({u\over r}\right)\right)^{-1/2}
$$
our state reproduces the boundary state of de Alwis after integration over $K$'s and ${\cal K}$'s with Gaussian weights\cite{SS2}.

An important feature of the state $|A,\Theta,\rho,\eta\rangle$ is that it is an eigenstate of the local coordinate of the strings at the boundary. Namely we can derive
\begin{equation}
\left(X^\zeta(\sigma)+i\epsilon\psi^\zeta(\sigma)\right)|A,\Theta,\rho,\eta\rangle
=\left(A(\sigma)+i\epsilon\Theta(\sigma)\right)|A,\Theta,\rho,\eta\rangle
\label{eigenvalue equation}
\end{equation}
when $\zeta\rho=1,\ \eta=-1$. We notice that the bosonic part of the operator $\hat W$ in (\ref{hat}) can be interpreted as a Wilson loop factor if we write it as $\exp\left[(i/4\pi)\oint A(\sigma)dX(\sigma)\right]$. From this point of view the equation (\ref{eigenvalue equation}) shows a relation of the string coordinate on a D-brane with a gauge field. 

On the other hand if we further specify the momentum distribution function such as $K(\sigma)=2\pi\sum_jk_j\delta(\sigma-\sigma_j)$ we find that the soliton coordinate $\xi(\sigma)$ of (\ref{Psi_0}) appears as the eigenvalue of the bosonic coordinate $X^\zeta(\sigma)$ attached on the D-brane. This interpretation sounds completely different from the previous gauge field interpretation. In order to reach to a consistent interpretation we must proceed further our consideration of the correspondence of the Sato-Wilson formulation and the string theory.

It is well known that the Sato-Wilson formalism can be extended straightforwardly to a matrix form. Since we have established the correspondence between the SW formulation and the closed string theory as given by (\ref{Psi(z)=string correlator}) we can naturally generalize the string theory to a matrix form correspondingly. This generalization certainly guarantees integrability of the string theory thus obtained.

In the KP theory the generalization is done by substituting a diagonal matrix in the place of soliton coordinate $\xi(\sigma)$ and general matrix in the place of the pseudo differential operator $W$. In our string formulation, it amounts to replace $k_j$'s and $\kappa_j$'s in (\ref{hat V}) by their corresponding $r\times r$ diagonal matrices $\bmm{k}_j$'s and $\bmm{\kappa}_j$'s. Similarly we generalize the gauge fields $A(\sigma)$ and $\Theta(\sigma)$ by $r\times r$ matrices $\bmm{A}(\sigma)$ and $\bmm{\Theta}(\sigma)$, so that we obtain the matrix generalization of operators:
\begin{eqnarray}
{\hat{\bmm{V}}}^\zeta&=&\exp\left[{i\sqrt2}\sum_j\left({\bmm{k}}_jX^\zeta_<(\sigma_j)-i{\bmm{\kappa}}_j\psi^\zeta_<(\sigma_j)\right)\right],
\label{matrix hat V}\\
{\hat{\bmm{W}}}^\zeta
&=&\exp\left[{i\over 4\pi}\oint\bmm{A}(\sigma)dX^\zeta_>(\sigma)-{1\over 4\pi}\int_0^{2\pi} \bmm{\Theta}(\sigma)\psi^\zeta_>(\sigma)e^{-i\sigma/2}d\sigma\right]
\label{matrix hat}
\end{eqnarray}

The matrix generalization of the gauge fields $A$ and $\Theta$ leads us to a nonabelian gauge theory. In the modern string theory the extension of an abelian gauge theory to nonabelian theory is provided by introducing coincident D-branes. It can be understood naturally within our framework. Namely we consider the boundary state $|\eta,\rho\rangle$ as a state of $r$ coincident D-branes. Denote the $i$th D-brane by $|\eta,\rho,i\rangle$. The state $|\bmm{A},\bmm{\Theta},\eta,\rho,i\rangle$ is obtained simply by a gauge rotation ${\bmm{\hat W}}$:
$$
|\bmm{A},\bmm{\Theta},\eta,\rho,i\rangle={\bmm{\hat W}}|\eta,\rho,i\rangle=\sum_{j=1}^r\hat{\bmm{W}}_{ij}|\eta,\rho,j\rangle.
$$

Now it is interesting to ask what is the expectation value of the string coordinate on the multi coincident D-branes. The same question is that how the eigenvalue formulae (\ref{eigenvalue equation}) will be generalized. After a little manipulation we find
$$
{\langle\bmm{A},\bmm{\Theta},\eta,\rho,i|\left(X^\zeta(\sigma)+i\epsilon\psi^\zeta(\sigma)\right)|\bmm{A},\bmm{\Theta},\eta,\rho,j\rangle
\over
\langle\eta,\rho,i|\eta,\rho,j\rangle}
=\bmm{A}_{ij}(\sigma)+i\epsilon\Theta_{ij}(\sigma).
$$
This formula tells us that the string coordinate on the coincident D-brane turns to appear as a matrix when we evaluate its expectation values.

\acknowledgements
{The authors would like to thank Mr. Ryuichi Sato for discussions. }

%%%%%%%%%%%%%%%%%%%%%%%%%%%%%%%%%%%%%%%%%%%%%%%%%

\end{document}